\documentclass[aps,prl,amsfonts,amssymb,amsmath,twocolumn,superscriptaddress,floatfix]{revtex4}

\usepackage[dvips]{graphics}
\usepackage{graphicx}

\def\cN{{\cal N}}
\def\cNf{{{\cal N}_f}}

\def\Tr#1{\mbox{Tr}\left[#1\right]}

\begin{document}

%\preprint{}

\title{Low-Temperature Thermal Conductivity of Superconductors With Gap Nodes}

\author{Tomas L\"ofwander}
\affiliation{Institut f\"ur Theoretische Festk\"orperphysik,
Universit\"at Karlsruhe, 76128 Karlsruhe, Germany}
\thanks{Present address.}
\affiliation{Department of Physics \& Astronomy, Northwestern University,
Evanston, Illinois 60208, USA}

\author{Mikael Fogelstr\"om}
%\thanks{}
\affiliation{Applied Quantum Physics Laboratory, MC2,
Chalmers University of Technology,
S-412 96 G\"oteborg, Sweden}

\date{March 16, 2005}

\begin{abstract}
  We report a detailed analytic and numerical study of electronic
  thermal conductivity in $d$-wave superconductors. We compare theory
  of the cross over at low temperatures from $T$-dependence to
  $T^3$-dependence for increasing temperature with recent experiments
  on $\mathrm{YBa_2Cu_3O_7}$ in zero magnetic field for $T\in
  [0.04K,0.4K]$ by Hill {\it et al.}, Phys. Rev. Lett. {\bf 92},
  027001 (2004). Transport theory, including impurity scattering and
  inelastic scattering within strong coupling superconductivity, can
  consistently fit the temperature dependence of the data in the lower
  half of the temperature regime. We discuss the conditions under
  which we expect power-law dependences over wide temperature
  intervals.
\end{abstract}

\maketitle

Low-temperature measurements of electronic transport properties have
given a wealth of information about nodal quasiparticles deep in the
superconducting phase of unconventional superconductors
\cite{bonn92,tai97,cor00,chi00,nak01,tak02,pro02,and02,suz02,sut03,bel04,hill04,sun04}.
The ultra-low temperature regime is of great interest because response
functions such as the zero-frequency charge conductivity, thermal
conductivity, and thermoelectric response function, are believed to be
limited by elastic impurity scattering and obey power laws
\cite{lee93,sun95,graf96,LF04}. Thermal conductivity is of particular
importance in this context, since theory predicts universality in the
sense that the low-temperature asymptotic does not depend on the
properties of the impurity potential \cite{durst00}. For higher
temperatures, of the order of the low-energy impurity band width
$\gamma$ and higher, thermal conductivity is not universal and is
sensitive to the details of impurity scattering \cite{graf96}.

The universal character of the $T\rightarrow 0$ thermal conductance
was studied experimentally in great detail for several different
cuprate materials for a wide range of doping levels
\cite{tai97,chi00,nak01,tak02,suz02,sut03,bel04}. Measurements of how
the universal limit is approached as temperature is lowered was only
recently reported by Hill {\it et al} \cite{hill04} for
$\mathrm{YBa_2Cu_3O_7}$. After subtracting a phonon contribution
$\propto T^3$, they found an electronic contribution of the form
$\kappa_{el}/T=(\kappa_{0}/T)(1+ k T^2)$, with
$\kappa_{0}/T=0.16\,{\rm mW/(K^2cm)}$ and the coefficient $k=19.2
K^{-2}$. This form is consistent with a Sommerfeld-type expansion of
$\kappa_{el}(T)$ (\cite{graf96} and below). However, the large value
of $k$ implies an unrealistically clean sample if impurity scattering
is in the unitary limit. On the other hand, for Born limit scattering,
the concentration required to fit the large slope is unrealistically
large and should lead to a severe reduction of the critical
temperature which is not seen. Their plausible conclusion was
therefore that scattering might be in between the unitary and Born
limits.

In this paper we present results for thermal conductivity including
effects of elastic impurity scattering and inelastic electron-boson
scattering. We focus on how the universal limit is approached and make
a detailed comparison with the experimental results of Hill {\it et
al} \cite{hill04}. Our results can be summarized as follows. (1) The
main effect of inelastic scattering is an effective mass-dependence of
the universal (with respect to impurity scattering) $T\rightarrow 0$
thermal conductance \cite{sch98}. The coefficient $k$ is however
effective mass independent. (2) We can fit the data \cite{hill04} in
the lower half of the $T-$interval with a best fit for a phase shift
slightly below $90^{\rm o}$. (3) The discrepancy at higher $T$ is
related to the split of the low-energy impurity band, which makes it
hard to observe power laws over wide $T-$intervals in ultra-clean
materials.

\begin{figure}[t]
\includegraphics[width=8cm]{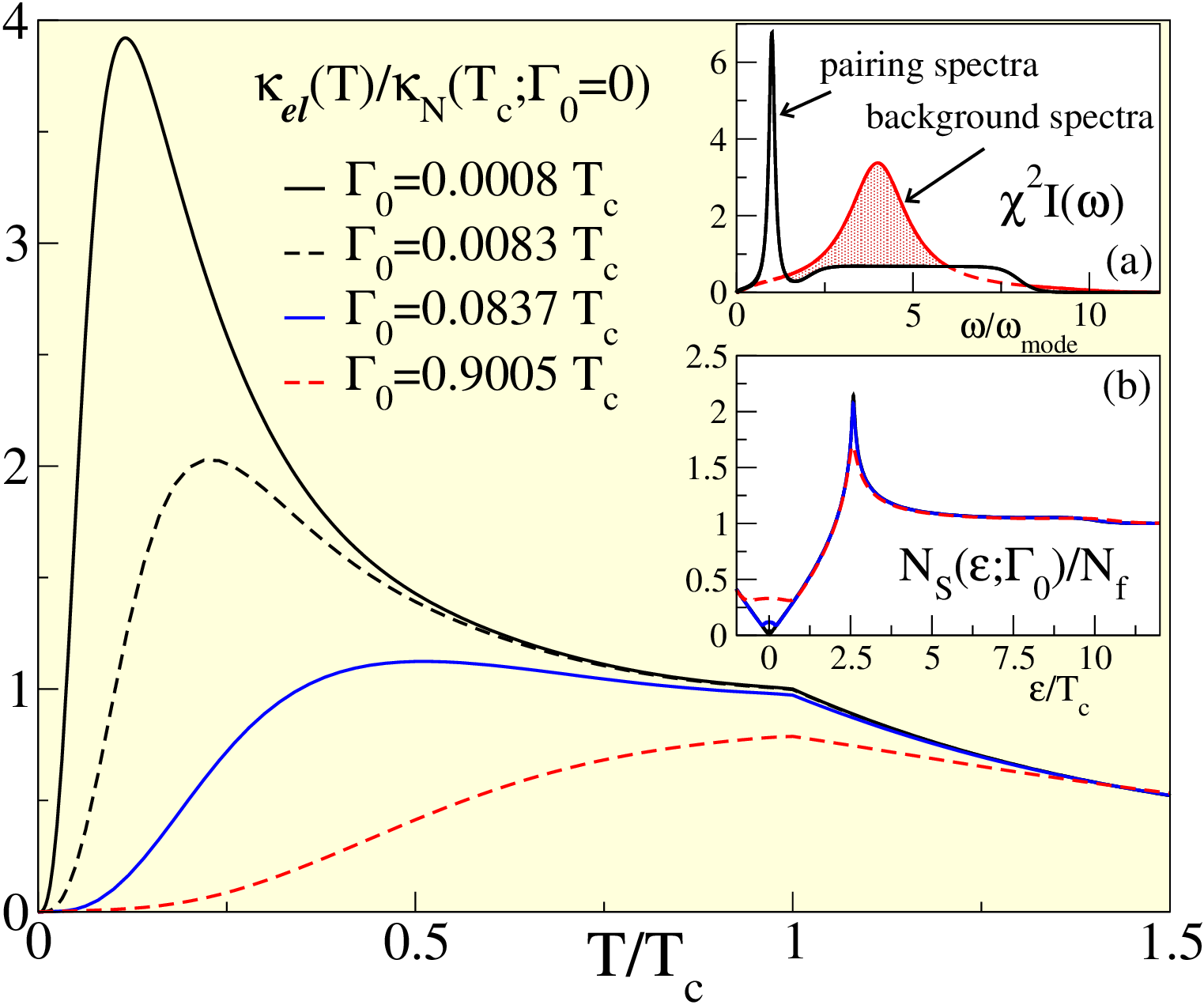}
\caption{\footnotesize Temperature dependence of the electronic
  contribution to the thermal conductivity for elastic scattering
  rates as indicated, for a phase shift $\delta_0=89.5^{\rm
  o}$. Insert (a): the two electron-boson coupling spectra used in the
  calculations. Insert (b): single-particle density of states in the
  superconducting state at $T=0$.}
\label{fig:Fig1}
\end{figure}

The thermal conductivity $\kappa(T)$ of a superconductor has an
electronic contribution $\kappa_{el}(T)$, a phononic part
$\kappa_{ph}(T)$, and possibly contributions from other existing
excitations. We will only consider $\kappa_{el}(T)$. To quantify the
effects of inelastic scattering on the same footing as elastic
impurity scattering we make a self-consistent strong-coupling
calculation using quasi-classical Eliashberg theory \cite{RS}
including impurity scattering in the $\hat t-$matrix approximation. We
consider a two-spectra model for electron-boson interactions that
generate both d-wave superconductivity and an inelastic scattering
self energy. One spectrum gives coupling in the s-wave and d-wave
channels, but is attractive only in the d-wave. Hence, this part is
responsible for pairing but also contributes to inelastic
scattering. The second spectrum couples only in the s-wave channel
(repulsively) and gives an incoherent background
scattering. Even though the model is general we choose parameters that
generates a d-wave superconductor with properties close to that of a
weakly coupled superconductor, with a spectroscopic gap
$\Delta_0=2.59\, T_c$ ($T_c=0.13 \omega_{\sf mode}$) \cite{note}. The
temperature dependence $\kappa_{el}(T)$ is displayed in
Fig.~\ref{fig:Fig1} for several impurity concentrations $n_{\sf imp}$
($\Gamma_0=n_{\sf imp}/\pi\cNf$) and near-resonant impurity scattering
(phase shift $\delta_0=89.5^{\rm o}$). We show in panels (a) and (b)
the electron-boson coupling functions, $\chi^2I(\omega)$, and the
resulting density of states, $N_S(\epsilon)$. The imaginary part of
the inelastic self energy gives a $T-$dependent contribution to the
scattering rate , $\Gamma_{\sf in}(T)$ \cite{msv88,hir93,hir96}, while
the contribution from elastic impurity scattering,
$\Gamma=\Gamma_0\sin^2\delta_0$, is $T-$independent. The relative
importance of the two contributions leads to a cross-over temperature
$T^*$. Impurity scattering dominates in the low$-T$ region, while
inelastic scattering dominates in the high$-T$ region. The cross-over
is reflected in $\kappa_{el}(T)$ as the peak at $T\sim T^*$ in
Fig.~\ref{fig:Fig1} \cite{hir96,sch98}.

Although the imaginary part of the inelastic self-energy
$\hat\Sigma^R_{\sf in}$ is small for $T\ll T^*$, the real part remains
important. By using general symmetry relations we can write down the
renormalization of the energy from both elastic and inelastic
scattering as
\begin{equation}
\tilde\epsilon^R
= \epsilon - \frac{1}{2} \Tr{\hat\tau_3 \hat\Sigma^R(\epsilon)}
= i\gamma + (1+\lambda)\epsilon + {\cal O}\left[\epsilon^2\right].
\label{eq:expansion}
\end{equation}
Here is $\gamma$ solely due to impurity scattering. The slope of the
real part of $\tilde\epsilon^R(\epsilon)$ has contributions from both
impurity scattering and inelastic scattering, $\lambda=\lambda_{\sf
imp}+\lambda_{\sf in}$. Because $\Re \Sigma^R_{\sf in}(\epsilon)$ is
linear in energy up to the mode-energy, $\omega_{\sf mode}>\Delta_0\gg
T^*$, an effective theory can be made for $T \ll T^*$ by dividing the
transport equation by $1+\lambda_{\sf in}$ evaluated in the
superconducting phase. The inelastic part is then interpreted as an
effective mass via $m^*/m=1+\lambda_{\sf in}$.  The contribution from
impurity scattering on the other hand, has a non-trivial energy
dependence for $\epsilon \lesssim \Delta_0$, and can not be divided
out. Within the effective theory, we use Eq.~(\ref{eq:expansion}) and
the method of Graf {\it et al} \cite{graf96} to write down the low$-T$
form
\begin{equation}
\frac{\kappa_{el}(T)}{T} =
\frac{\pi^2}{3}\frac{2\cNf v_f^2}{\pi \mu \Delta_0}
\left[1 + \frac{7\pi^2}{15}\frac{\tilde a^2T^2}{\gamma^2}\right]
+ {\cal O} \left[\left(\frac{T}{\gamma}\right)^4\right].
\label{eq:exp}\end{equation}
Here are $\cNf$ and $v_f$ the {\em effective} density of states at the
Fermi level and the {\em effective} Fermi velocity, while $\Delta_0$
is the spectroscopic gap and $\mu$ is the opening rate of the gap
function at the node. We now find that the remaining effect at $T\ll
T^*$ of inelastic scattering within strong-coupling superconductivity
is a modification of the $T\rightarrow 0$ asymptotic of the thermal
conductance. When we write explicitly, $\cN_f\rightarrow
\cN_f^*=\cN_f^0(1+\lambda_{\sf in})$, and $v_f\rightarrow
v_f^*=v_f^0/(1+\lambda_{\sf in})$, one factor $1+\lambda_{\sf in}$
remains in the denominator. Within the bare theory, this result can be
traced back to that the spectroscopic gap $\Delta_0$ in the
weak-coupling limit is replaced by the strong coupling off-diagonal
function $W(0)$ \cite{sch98}. In our model $W(\epsilon)$ is a weak
function of energy and to a good approximation
$W(0)\approx\Delta_0(1+\lambda_{\sf in})$. This effective mass
dependence is in sharp contrast to the normal state, in which case
$\kappa_{el}^N(T)=T(\pi^2/3)\cN_f v_f^2/2 \Gamma$ is independent of
$m^*$, because of canceling factors of $(1+\lambda_{\sf in})$ in
$N_f, v_f^2$ and $\Gamma$ in accordance with general strong-coupling
results for transport coefficients \cite{pra64}. In this sense,
transport properties of quasiparticles in the low-energy nodal
impurity band is quite different from quasiparticles in the normal
state.

Effective-mass factors from $\tilde a$ and $\gamma$ cancel in the
coefficient $k=\frac{7\pi^2}{15}(\tilde a/\gamma)^2$ in
Eq.~(\ref{eq:exp}) and it is only dependent on properties of the
impurity potential. In addition to the electron-hole symmetric part
above, also the electron-hole asymmetric part of the impurity
self-energy enters \cite{monien87b}. Only the imaginary part
\begin{equation}
\frac{1}{2}\Tr{\Im \hat\Sigma^R_{\sf imp}(\epsilon)} = c\,\epsilon
+ {\cal O} \left[\epsilon^3\right].
\label{eq:unitpart}
\end{equation}
contributes to the conductivity through $\tilde a^2=(1+\lambda_{\sf
imp})^2+2c^2$ \cite{LF04}. The presence and importance of $c$
represents the fact that electron-like and hole-like quasiparticles
have different transport lifetimes above and below the Fermi level,
except in the strict unitary and Born limits for which it vanishes.

\begin{figure*}[t]
\includegraphics[width=17cm]{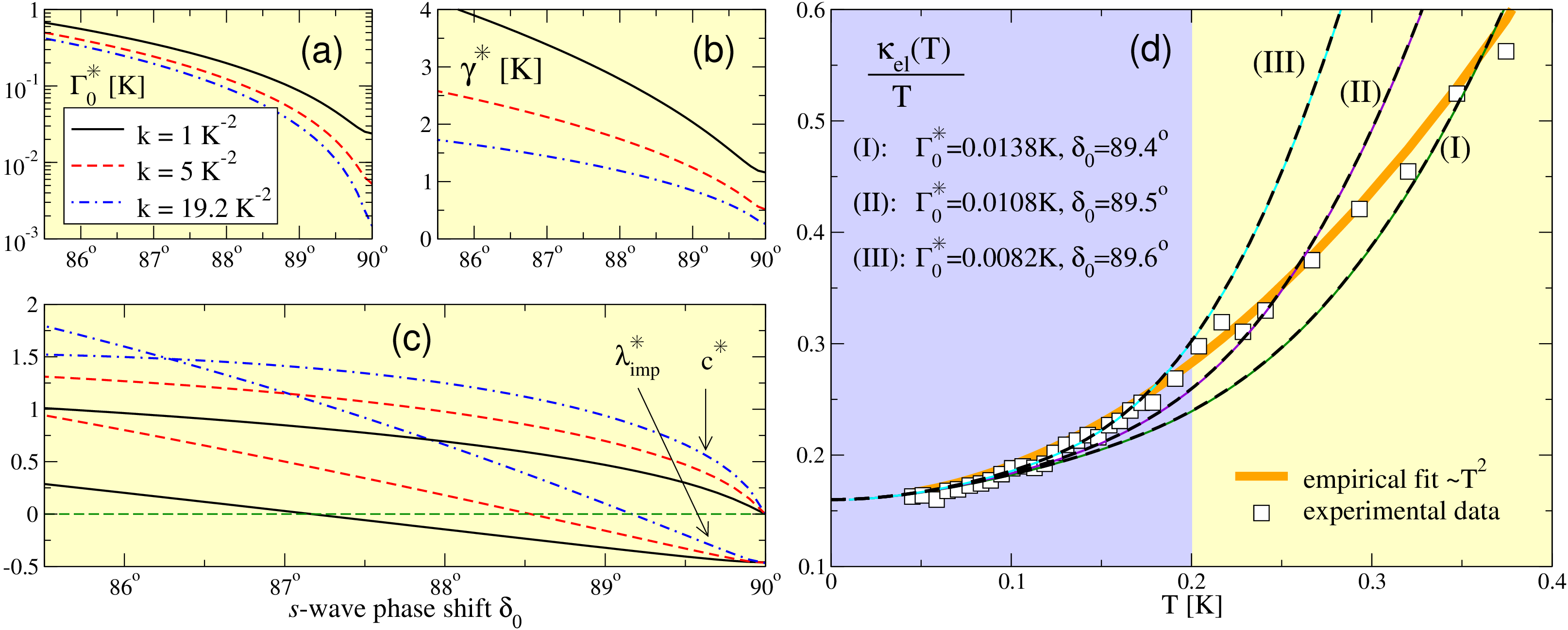}
\caption{\footnotesize (a)-(c) The low-temperature parameters in
  Eqs.~(\ref{eq:expansion}) and (\ref{eq:unitpart}) for three values
  of the $T^2$-coefficient in Eq.~(\ref{eq:exp}) as function of
  scattering phase shift. (d) Three sets of parameters that represents
  our best fits to the data of Ref.~\cite{hill04}. The blue shaded
  region indicates the temperature range where a good fit can be
  made. The dashed curves are from a strong-coupling calculation while
  the underlying solid lines are generated by the effective
  theory. Curve (II) is the low$-T$ part of the black solid line in
  Fig.~\ref{fig:Fig1}. In all cases $k=19.2K^{-2}$. Note that the
  material appears cleaner in the effective theory as
  $\Gamma_0^*=\Gamma_0/(1+\lambda_{\sf in})$ with $\Gamma_0=n_{\sf
  imp}/\pi\cNf$. We normalized $\kappa_{el}(T)/T$ to the $T\rightarrow
  0$ extrapolated value $0.16\,{\rm mW/(K^2 cm)}$. The energy scale is
  $T_c=90K$.}
\label{fig:fit}
\end{figure*}
We now discuss the experiment \cite{hill04} in terms of the presented
effective theory and its low$-T$ expansion in Eq.~(\ref{eq:exp}) with
the parameters $\{\gamma,\lambda_{\sf imp},c\}$ computed
self-consistently \cite{LF04}. Starting for simplicity with $s$-wave
impurity scattering, we present in Fig.~\ref{fig:fit}(d) several
combinations of impurity concentrations and phase shifts that give
good fits to the data in the lower half of the $T-$interval. In fact,
there is a broad range in parameter space for which the slope
$k=19.2K^{-2}$ can be reproduced [blue dash-dotted lines in
Fig.~\ref{fig:fit}(a)-(c)]. Similarly, we can find many combinations
yielding other values of $k$ (black and red lines). However, not all
combinations of parameters produce a consistent fit, in the sense that
the expansion Eq.~(\ref{eq:exp}) represents over a wide temperature
range the full numerically computed thermal conductivity shown in
Fig.~\ref{fig:fit}(d). The deviation from true $T^2$-behavior is also
seen in the figure: the three curves start at low $T$ with the same
coefficient, $k=19.2K^{-2}$, but deviate at higher $T$. This is also
the reason for the discrepancy between theory and experiments in the
upper half of the $T-$interval. The part in parameter space for which
the Sommerfeld expansion is applicable can be broadly defined as phase
shifts for which $\lambda_{\sf imp}$ is negative.

\begin{figure}[t]
\includegraphics[width=8cm]{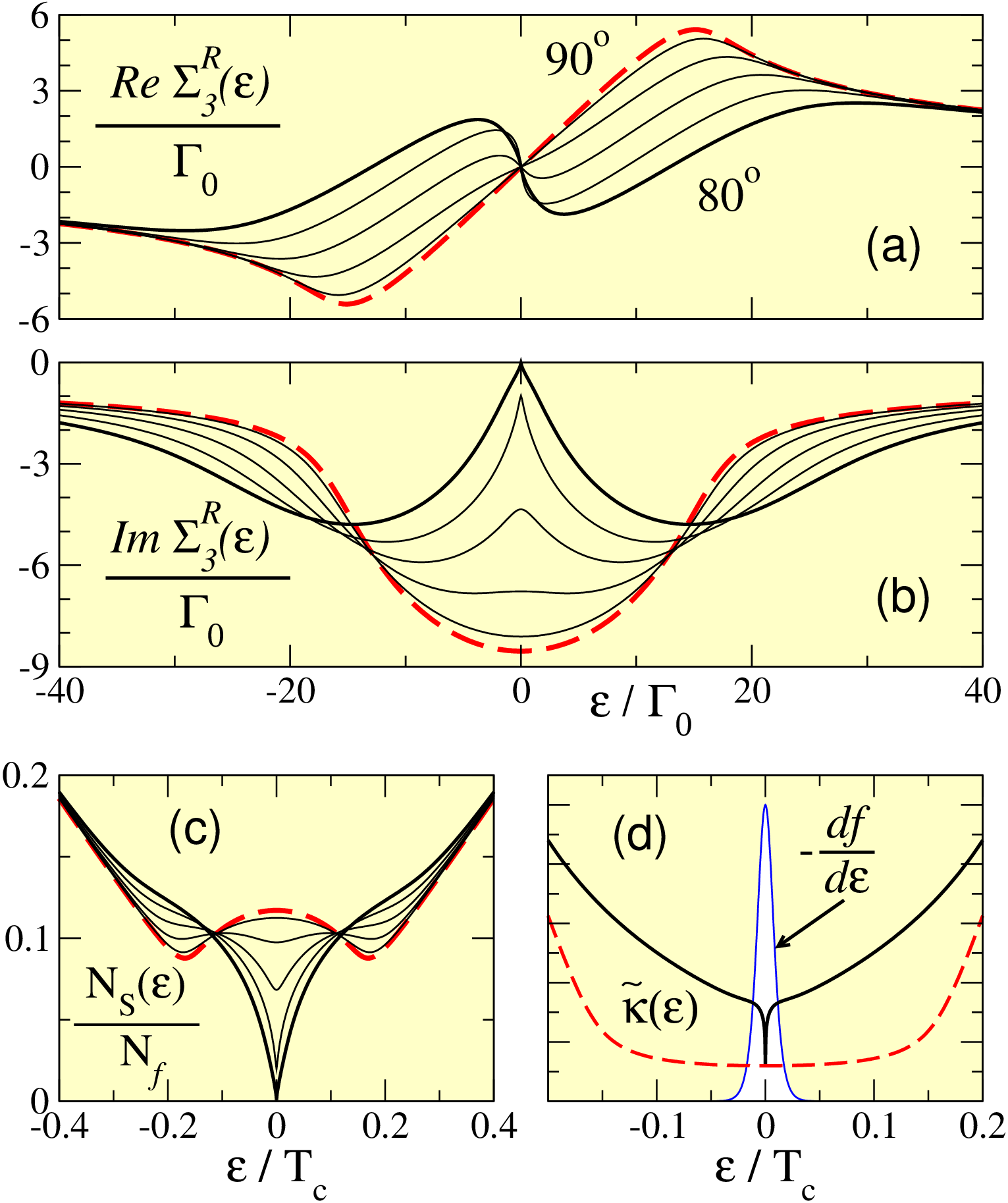}
\caption{\footnotesize (a)-(b) The impurity scattering self-energy and (c) the
  density of states for $\Gamma_0=0.01T_c$ and phase shifts ranging
  from $90^{\sf o}$ (red dashed line) to $80^{\sf o}$ (thick black
  line) in steps of two degrees. (d) The kernel
  $\tilde\kappa(\epsilon)$ of the thermal conductivity contains
  structures at low energies for phase shifts below $90^{\sf o}$
  because of the split impurity bands seen in (a)-(c). Also shown is
  the derivative of the Fermi function which has a peak at the Fermi
  level of width $\sim T$. Here is $T=0.005T_c=0.45K$ for $T_c=90K$.}
\label{fig:sR30}
\end{figure}

Let us discuss in more detail why Eq~(\ref{eq:exp}) works well only in
a restricted low-$T$ interval. Recall that the thermal conductivity
can be written as
%
%\begin{equation}
$
\kappa_{el}(T)=\int_{-\infty}^{\infty} d\epsilon K(\epsilon)
\left( -\frac{\partial f}{\partial\epsilon}\right),
$
%\end{equation}
%
where $f$ is the Fermi function. The derivative of the Fermi function
is sharply peaked at the Fermi level in an interval of order $T$,
which leads us to expand all quantities around $\epsilon=0$, as in
Eqs.~(\ref{eq:expansion}) and (\ref{eq:unitpart}), and do a Sommerfeld
expansion
%
%\begin{equation}
$
\kappa_{el}(T)=K(0) + \sum_{n=1}^{\infty}
\left.\frac{d^{2n}K(\epsilon)}{d\epsilon^{2n}}\right|_{\epsilon=0}
a_n T^{2n},
$
%\end{equation}
%
where the for us relevant coefficients $a_1=\pi^2/6$ and
$a_2=7\pi^4/360$. However this scheme works well only if the function
$K(\epsilon)$ is slowly varying on the scale $T$. In our case we can
write $K(\epsilon)=\epsilon^2\tilde\kappa(\epsilon)$ and study the
energy dependence of $\tilde\kappa(\epsilon)$, see
Fig.~\ref{fig:sR30}(d). The fine structure at low energies for phase
shifts below $90^{\rm o}$, can intuitively be understood by studying
the phase shift dependence of the self-energy. In a $d$-wave
superconductor with a single impurity scattering strongly (near the
unitary limit) there is a resonance in the ${\hat t}-$matrix at
\cite{bal95} $\epsilon_{res}\approx \pm\pi\cot\delta_0 /
[2\ln(8/\pi\cot\delta_0)]$ where the sign corresponds to the electron
and hole sectors. The width of the resonance is smaller than the
resonance energy by a factor $1/\ln(8/\pi\cot\delta_0)\ll 1$. Within
transport theory we employ an impurity averaging technique \cite{AGD}
to obtain the self-energy for a material with a small concentration of
randomly distributed impurities. As a result of the configuration
average, the single impurity resonance is broadened into an impurity
band of width $\gamma_b$, centered around an energy slightly shifted
from the single impurity resonance
$\epsilon_{res}\rightarrow\epsilon_b$. The impurity bands are shown in
Fig.~\ref{fig:sR30}(a)-(c). For clean systems with a phase shift
deviating substantially from $90^{\rm o}$, the resulting bands are
split into two parts symmetrically positioned above and below the
Fermi level. The critical point is when the slope at zero energy of
the real part of the self energy changes sign, i.e. when $\lambda_{\sf
imp}=-\Re\Sigma^R_3(0)$ turns positive. The structures in the
self-energy lead to structures also in the response functions at low
temperatures on the scale $T\sim\epsilon_b\ll\gamma_b$, see
Fig.~\ref{fig:sR30}(d), and the applicability of Eq.~(\ref{eq:exp}) is
drastically reduced in temperature by the new constraint
$T\ll\epsilon_b$.

In a more realistic model, impurity scattering is anisotropic. The
main technical complication in this case is the necessity to take care
of vertex corrections. They are small in the limit $T\ll \gamma$
\cite{durst00}, but are important for higher temperatures. In
particular they contribute to the coefficient $\tilde a$ in
Eq.~(\ref{eq:exp}). For definiteness we have included $p$-wave and
$d$-wave scattering channels and considered arbitrary phase shifts
\cite{unpub}. In general we obtain a pair of resonances, one
electron-like and one hole-like (symmetrically positioned on each side
of the Fermi level), for each phase shift. This fact was also noted
recently in Ref.~\cite{nunner04}. The main conclusion relevant in the
present discussion is that the impurity self-energy aquires more
structures (related to the new resonances) and the range of
applicability of Eq.~(\ref{eq:exp}) is not improved.

In summary, let us list our results and make a few remarks. (1)
Inelastic scattering contributes to the $T\rightarrow 0$ asymptotic
(i.e. $\kappa_0\propto T$), but plays no role for the $T^3$
coefficient (i.e. $k$). (2) The low$-T$ half of the experimental data
can be consistently fitted with parameters representing a wide range
of impurity concentrations and phase shifts. The best fit to the large
coefficient $k=19.2K^{-2}$ is obtained for a rather clean material
with a scattering phase shift slightly below the unitary limit
($\delta_0<90^{\rm o}$). (3) However, we find deviations in the
high$-T$ region. The reason is that the expansion Eq.~(\ref{eq:exp})
has a limited range of applicability, smaller than the temperature
range in which the $T^3$ term was found experimentally to
dominate. Indeed, in an earlier discussion of the data \cite{hill04}
in Ref.~\cite{kim03}, model parameters were chosen so that the
expansion Eq.~(\ref{eq:exp}) was inapplicable. In their work the
unit-term Eq.~(\ref{eq:unitpart}) was neglected, which changed their
results by about a factor two. We conclude that a clear cross over
between $T-$dependence to $T^3$-dependence, with the respective
power laws holding in wide $T-$intervals, is most easily observable in
dirty to intermediately clean, but not in ultra-clean,
materials. Finally, an alternative way of extracting information about
impurity scattering from transport properties, would be to study the
thermoelectric effect \cite{LF04}. In this case, the non-universality
of the $T\rightarrow 0$ asymptotic can be used. This quantity, the
analogue of $\kappa_0$, should be easier to measure and interpret than
the $T^3-$part of $\kappa_{el}$.

\begin{acknowledgments}
  We gratefully acknowledge financial support from the Swedish
  Research Council (M.F.), the Wenner-Gren foundations (T.L.), and the
  EC under the Spintronics Network RTN2-2001-00440 (T.L.).
\end{acknowledgments}

%\bibliography{thermoLowT}

\begin{thebibliography}{29}
\bibitem[{not()}]{note}
\bibinfo{note}{The electron-boson coupling strength is $\lambda_{\sf in}=6.0$,
  with 48.5\% contribution from $\lambda_{\sf d-wave}$ and 51.5\% from
  $\lambda_{\sf incoh}$.}
\bibitem[{unp()}]{unpub}
\bibinfo{howpublished}{The details of this calculaton will be presented
  elsewhere.}
\end{thebibliography}

\end{document}